\documentclass[twocolumn,showpacs,
  aps,superscriptaddress,
  prd,notitlepage,showkeys,
  nofootinbib,10pt]{revtex4-2}
\usepackage{blindtext}
\usepackage[normalem]{ulem}
\usepackage{amssymb}
\usepackage{amsmath}
\usepackage{graphicx,subfigure}
\usepackage{dcolumn}
\usepackage[colorlinks,
            linkcolor=blue!50!black,
            citecolor=green!50!black,
            urlcolor=blue!70!black]{hyperref}
\usepackage{color,units}
\usepackage[dvipsnames]{xcolor} 
\usepackage{lineno}
\usepackage{xspace}
\usepackage{longtable} 
\usepackage{float}  
\usepackage{aas_macros}
\usepackage{amsfonts,wasysym,epsfig,verbatim,subfigure,bm,mathrsfs,lipsum}
\usepackage{comment}
\usepackage{braket}
\usepackage{physics}
\usepackage{hyperref}
\usepackage{orcidlink}
\usepackage{cuted}

\DeclareRobustCommand{\varlambda}{\text{\usefont{OML}{txmi}{m}{it}\symbol{"15}}}

\begin{document}

\title{Bose-Einstein Condensate Dark Matter in the Core of Neutron Stars: Implications for Gravitational-wave Observations}

\author{Samanwaya Mukherjee\orcidlink{0000-0002-9055-5784}}
\email{samanwaya.physics@gmail.com}
\affiliation{International Centre for Theoretical Sciences, Tata Institute of Fundamental Research, Bengaluru - 560089, India}
\affiliation{Inter-University Centre for Astronomy and Astrophysics (IUCAA), Post Bag 4, Ganeshkhind, Pune 411007, India}

\author{P.~S.~Aswathi\orcidlink{0009-0008-1458-3338}}
\email{aswathipampurayam@gmail.com}
\affiliation{Inter-University Centre for Astronomy and Astrophysics (IUCAA), Post Bag 4, Ganeshkhind, Pune 411007, India}
\affiliation{OzGrav-ANU, Centre for Gravitational Astrophysics, College of Science, The Australian National University, Australian Capital Territory 2601, Australia}

\author{Chiranjeeb Singha\orcidlink{0000-0003-0441-318X}}
\email{chiranjeeb.singha@iucaa.in}
\affiliation{Inter-University Centre for Astronomy and Astrophysics (IUCAA), Post Bag 4, Ganeshkhind, Pune 411007, India}

\author{Apratim Ganguly\orcidlink{0000-0001-7394-0755}}
\email{apratim@iucaa.in}
\affiliation{Inter-University Centre for Astronomy and Astrophysics (IUCAA), Post Bag 4, Ganeshkhind, Pune 411007, India}

\newcommand{\sm}[1]{\textsf{\color{red}[{ #1}]}}
\newcommand{\PSA}[1]{\textsf{\color{magenta}[{ #1}]}}

\begin{abstract}

We investigate neutron stars admixed with dark matter (DM) in the form of a finite-temperature Bose–Einstein condensate (BEC) within a general relativistic two-fluid framework in which the nuclear and dark components interact only gravitationally. 
Using realistic nuclear matter equations of state (EOS), APR4, MPA1, and SLy, we construct equilibrium configurations and compute mass–radius relations, tidal Love numbers, and dimensionless tidal deformabilities. 
We quantify how the presence of a BEC dark component modifies the mass–$\Lambda$ relation relevant for gravitational wave observations, finding that increasing the DM mass fraction generically reduces the maximum mass, radius, and tidal deformability of neutron stars. 
By comparing theoretical mass–$\Lambda$ curves with EOS-insensitive posteriors from GW170817, we evaluate, in a conditional sense, the dark matter fractions that would align a given nuclear EOS with the observed tidal constraints; for example, under the assumption that APR4 describes nuclear matter and that the GW170817 components were dark-matter admixed neutron stars, our study favors dark matter fractions of order a few percent, whereas stiffer EOSs require larger fractions to achieve comparable agreement. 
This interpretation assumes that inspiral waveforms are adequately characterized by tidal deformability and should therefore be regarded as structural rather than a direct detection of dark matter. 
We also examine finite-temperature effects in the BEC sector and find that, for moderate dark matter fractions, temperature has a negligible impact on the stability and tidal properties of admixed configurations. 
Our results demonstrate how even modest DM admixtures can influence neutron star structure and tidal observables, highlighting the importance of considering non-standard matter components in multimessenger constraints on dense matter.

\end{abstract}

\maketitle

\section{\label{sec:intro}Introduction}

Astrophysical and cosmological observations strongly indicate that most of the matter content of the Universe is in the form of dark matter~\cite{Planck:2018vyg, SDSS:2014iwm, Khlopov:2021xnw}. 
The evidence for DM arises entirely from its gravitational influence on visible matter, radiation, and large-scale structure. 
Despite various theoretical and experimental efforts, including particle accelerators \cite{CMS:2012ucb, Klasen:2015uma} and nuclear recoil experiments~\cite{ATLAS:2012ky}, the microscopic nature of DM remains unknown. 
Since all forms of matter gravitate, compact astrophysical objects provide a complementary avenue to probe possible DM contributions through their impact on macroscopic observables.

Neutron stars (NSs), the densest stellar remnants described by the standard model of physics, are particularly promising candidates for such investigations. 
Observation of massive pulsars exceeding $2M_\odot$~\cite{2013Sci...340..448A, Fonseca:2016tux, Demorest:2010bx, NANOGrav:2019jur} and radius measurements from the NICER mission~\cite{Riley:2019yda, Miller:2019cac, Riley:2021pdl, Miller:2021qha} have significantly constrained the equation of state (EOS) of cold dense nuclear matter. 
However, the composition of matter at supranuclear densities remains uncertain. 
If an additional dark component is present within neutron stars, it can modify their global properties such as the mass–radius relation, maximum mass, compactness, and tidal deformability. 
DM-admixed neutron stars (DANSs) have been studied extensively within the two-fluid framework where nuclear matter (NM) and DM interact primarily through gravity (see, e.g., \cite{Goldman:1989nd, Ellis:2018bkr, Panotopoulos:2017idn, Karkevandi:2021ygv, Thakur:2023aqm, Collier:2022cpr}). 
These models demonstrate that even modest dark matter fractions can alter the structural properties of NSs. 
However, a key challenge arises from degeneracies: changes in macroscopic observables may be attributed either to the stiffness of the nuclear EOS or to the presence of DM. 
Disentangling these effects is therefore nontrivial.

The detection of gravitational waves (GWs) from binary neutron star mergers, beginning with GW170817~\cite{LIGOScientific:2017vwq, LIGOScientific:2018cki}, has opened a new window to probe the internal structure of neutron stars. 
During the inspiral phase, tidal effects modify the gravitational waveform through the dimensionless tidal deformability parameter $\Lambda$, which depends sensitively on the stellar compactness \cite{Hinderer:2007mb}. 
Measurements of the reduced tidal deformability $\tilde{\Lambda}$, the directly measurable quantity in parameter estimation analyses from GW data, thus provide direct constraints on the 
relation between $\Lambda$ and the NS mass $M$. 
If DM is present within neutron stars, it modifies the stellar compactness and the Love number, thereby affecting $\Lambda(M)$~\cite{Dengler:2021qcq, Sen:2021wev, Collier:2022cpr, Thakur:2023aqm}. GW data may also help constrain the interaction cross sections of non-annihilating DM particles with nucleons~\cite{Bhattacharya:2023stq}.

In this work, we consider DM in the form of a self-gravitating Bose-Einstein condensate (BEC). 
Such models have been studied extensively in astrophysical contexts~\cite{Boehmer:2007um, Harko:2011xw, Chavanis_2011}, 
and their presence inside neutron stars has been investigated in the zero temperature limit~\cite{Ellis:2018bkr}. 
Here, we extend such studies by employing a finite-temperature BEC EOS~\cite{Gruber:2014mna} and examine its implications for NS structure and tidal deformability. 
This EOS was previously used to study rotating BEC stars at finite temperature~\cite{Aswathi:2023zzn}, and has also been applied in the context of modified gravity theories, such as combined Rastall-Rainbow gravity~\cite{Jyothilakshmi:2023cao}, highlighting its broader applicability in compact star modeling. 
We adopt a general relativistic two-fluid formalism to construct equilibrium configurations of DANSs using realistic nuclear EOSs (APR4~\cite{Akmal:1998cf}, MPA1~\cite{Muther:1987xaa}, and SLy~\cite{Douchin:2001sv}) consistent with current pulsar and GW constraints. 
We compute mass–radius relations, tidal Love numbers, and dimensionless tidal deformabilities across a range of DM fractions and temperatures.

A central astrophysical question concerns the amount of DM that a NS may realistically contain. Standard capture scenarios typically predict small DM mass fractions, often $\lesssim 10^{-3}$ even in dense environments~\cite{Ivanytskyi:2019wxd}. 
Larger fractions may require non-standard formation mechanisms or early-universe scenarios. 
In this study, we do not model the detailed accumulation history. 
Instead, we treat the DM mass fraction $f_{\rm DM}$ as a phenomenological structural parameter and investigate how its presence would modify the observable properties of NSs. 
Our goal is to quantify the structural and tidal effects under this hypothesis 
without adopting any specific formation channel.

Our interpretation of GW constraints focuses primarily on the mass–$\Lambda$ relation. 
We assume that dark matter modifies the waveform predominantly through its effect on the stellar structure and tidal deformability. 
Radius inferences, which rely on quasi-universal relations calibrated for ordinary neutron stars, are treated cautiously in the presence of non-standard matter components.

The paper is organized as follows. In Sec. \ref{fomal}, we describe the finite-temperature BEC equation of state and the two-fluid formalism, and outline the computation of tidal Love numbers. 
In Sec. \ref{result}, we discuss equilibrium configurations and analyze how dark matter affects the mass–radius and mass–$\Lambda$ relations in light of current observational constraints. 
Finally, in Sec. \ref{conclusion}, we summarize our findings and discuss possible extensions of this work. 

\textbf{\emph{Notations and Conventions}}:  
We adopt the metric signature $(-,+,+,+)$ and work in natural units with $G=c=\hbar=1$ throughout.

\section{Formalism} \label{fomal}

\subsection{Equations of state} \label{sec:eos}

\begin{figure}[!ht]\label{EoS}
\centering
\includegraphics[width=0.5\textwidth]{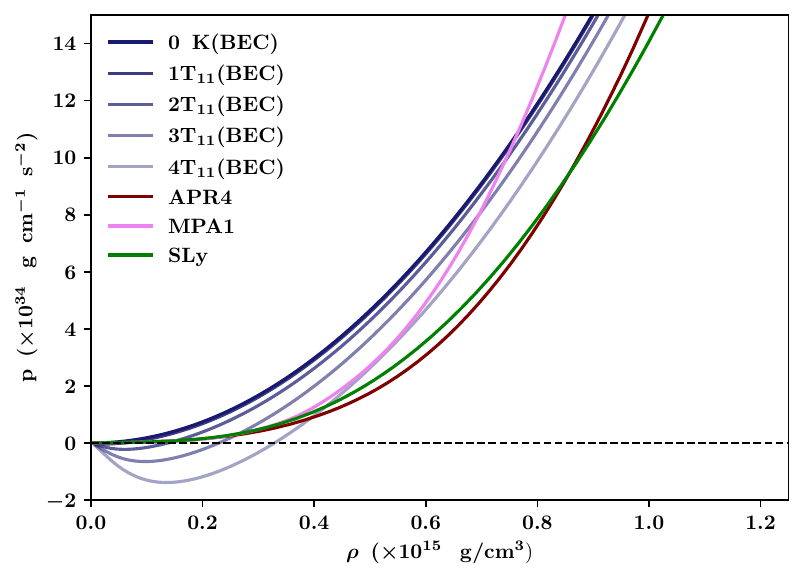}
\caption{Pressure–density relations for the finite-temperature BEC dark matter EOS (shown for $T_{11}=10^{11}\,$K) compared with the nuclear matter EOSs APR4, MPA1, and SLy.}
\end{figure}

We model DM as a self-gravitating BEC 
at finite temperature following Ref.~\cite{Gruber:2014mna}. 
The EOS relates the pressure $p$ and density $\rho$ as 
\begin{eqnarray} \label{eq:eos}
p(\rho) &=& \frac{g\rho^2}{2m^2} + \frac{2g\rho}{m \varlambda^3} \zeta_{3/2}\left[e^{-\beta g\rho/m}\right]
+ \frac{2}{\beta \varlambda^3} \zeta_{5/2}\left[e^{-\beta g\rho/m}\right]\nonumber\\&-& \frac{2}{\beta \varlambda^3} \zeta_{5/2}\left[1\right]~.
\end{eqnarray}
Here $g = 4\pi a/m$ characterizes the repulsive contact interaction with scattering length $a$ and boson mass $m$. 
The thermal de Broglie wavelength is $\varlambda =  \sqrt{(2\pi\beta)/m}$, where $\beta=1/(k_{\rm B} T)$.
The function $\zeta_{\mathcal{V}}$ denotes the polylogarithm 
\begin{equation}
    \zeta_{\mathcal{V}}[z] = \sum_{n=1}^{\infty}\frac{z^n}{n^{\mathcal{V}}}~.
\end{equation}
The first term in Eq.~\eqref{eq:eos} corresponds to the zero-temperature condensate contribution, while the remaining terms account for the thermal cloud. 
The constant term ensures that $p \to 0$ as $\rho \to 0$, maintaining physical consistency in the limit of low density. 
At very low densities, the Thomas–Fermi approximation, which neglects quantum pressure, leads to an unphysical negative pressure region, as seen in Fig.~\ref{EoS}. 
This behavior occurs near the stellar surface where the condensate density becomes small. 
Inclusion of the quantum pressure term would regularize this region and restore positivity of the pressure~\cite{Gruber:2014mna}. 
Since this regime contributes negligibly to the global stellar structure and tidal properties, we retain the Thomas–Fermi form throughout. 
The global properties of static and rotating systems utilizing this EOS have been extensively studied in Ref.~\cite{Aswathi:2023zzn}. 
We take the boson mass $m=2m_n$, where $m_n$ is the nucleon mass, motivated by the possibility that paired nucleons behave effectively as bosons~\cite{Chavanis:2011cz}, and adopt $a = 10~\mathrm{fm}$, which yields maximum masses compatible with the $2M_\odot$ constraint.
For NM, we consider realistic EOSs APR4~\cite{Akmal:1998cf}, MPA1~\cite{Muther:1987xaa}, and SLy~\cite{Douchin:2001sv}, all consistent with the observed existence of $2M_\odot$ pulsars and current tidal constraints from GW170817. 
We show their pressure-density relations along with the BEC EOS in Fig.~\ref{EoS}.
We restrict temperatures to $(1 \,\text{-}\, 4)\times10^{11}$ K, following Ref.~\cite{Gruber:2014mna}. 
Below this range thermal corrections are negligible, making a zero-temperature treatment adequate, while at higher temperatures numerical instabilities arise.

\subsection{Two-fluid formalism}

We model DANSs as two non-interacting fluids coupled only through gravity. The static spherically symmetric metric for a relativistic star is
\begin{align}
    \dd s^{2} 
   &=-e^{\nu(r)}\dd t^{2}+e^{\lambda(r)}\dd r^{2}+r^{2}\left(\dd\theta ^{2}+\sin ^{2}\theta \dd\phi ^{2}\right)\,,
\end{align}
with $\nu(r)$ and $\lambda(r)$ being the metric functions given by 
\begin{equation}
    e^{\lambda(r)} = \left[1 - \frac{2m(r)}{r}\right]^{-1}\,,
\end{equation}
and
\begin{equation}
    \dv{\nu(r)}{r}=2\frac{e^{\lambda(r)}}{r^2}{\left[m(r)+4\pi p(r)r^3\right]}\,,
\end{equation}
$m(r)$ being the total mass enclosed within a radius $r$.

The coupled Tolman–Oppenheimer–Volkoff equations~\cite{Tolman:1939jz, Oppenheimer:1939ne} for the two-fluid system are
\begin{subequations}\label{eq:tov}
  \begin{align}\label{TOV-2a}
    &\dv{p_{\rm NM}}{r} =-(p_{\rm NM} + \rho_{\rm NM})\frac{(m(r)+4\pi r^3 p)}{r^2(1-2m(r)/r)}~,\\
    &\dv{p_{\rm DM}}{r} =-(p_{\rm DM} + \rho_{\rm DM})\frac{(m(r)+4\pi r^3 p)}{r^2(1-2m(r)/r)}~,\\
    &\dv{m_{\rm NM}}{r} = 4\pi r^2 \rho_{\rm NM}~,\\ \label{TOV-2b}
    &\dv{m_{\rm DM}}{r} = 4\pi r^2 \rho_{\rm DM}~,
  \end{align}   
\end{subequations}
where $p_{\rm NM}$ ($p_{\rm DM}$) and $\rho_{\rm NM}$ ($\rho_{\rm DM}$) are the pressure and density of the nuclear matter (dark matter) component. 
Here, $p= p_{\rm NM} +p_{\rm DM}$ and $m(r)= m_{\rm NM}(r) + m_{\rm DM}(r)$ are the total pressure of the admixture and the total mass enclosed within the radius $r$, respectively. 

Given an EOS, these coupled differential equations can be numerically integrated out from the center to the surface of the star to obtain the equilibrium stellar model. 
By fixing the two central densities $\rho_{\rm NM}(r=0)$, $\rho_{\rm DM}(r=0)$ together with initial conditions at the center of the star ($m_{\rm NM}(r=0) = m_{\rm DM}(r=0) =0 $), equations \eqref{TOV-2a}-\eqref{TOV-2b} can be numerically integrated till the surface of the star to obtain the global properties. 
When DM is mostly localized within the NS and nuclear matter extends beyond the DM core, the integration is carried out till the pressure of the nuclear matter component falls to a negligible value. 
The radial distance of this point from the center corresponds to the observable radius ($R$) of the star. 
The mass ($M$) of the star, then, is the total mass contained within this radius $R$: $M=m(R)$. 
The dark matter fraction, which denotes the amount of DM present in the admixture relative to the total NS mass, is 
\begin{equation}
    f_{\rm DM} = \frac{m_{\rm DM}(R)}{M}\,,
\end{equation}
where $m_{\rm DM}(R)$ denotes the DM mass enclosed within the radius $R$.

\subsection{Tidal deformability and Love numbers}

When placed in an external quadrupolar tidal field $Q_{ij}$, a spherically symmetric star develops an induced quadrupole moment $\xi_{ij}$ in response. To linear order,
\begin{align}
    Q_{ij} = - \bm{\lambda} \xi_{ij}~.
\end{align}
$\bm{\lambda}$ is referred to as the (dimensionful) tidal deformability, which quantifies how much a star deforms under an external tidal field. 
Following the standard perturbative approach~\cite{Hinderer:2007mb}, we compute the $l=2$ polar Love number $k_2$ from the coupled perturbation equations, and obtain the dimensionless tidal deformability as
\begin{equation} \label{Lambda}
    \Lambda = \frac{2}{3}k_2^{\text{polar}} \Big(\frac{M}{R}\Big)^{-5}\,.
\end{equation}

\paragraph*{\textbf{Interpretation of tidal deformability in GW observations}:}
The reduced tidal deformability measured in binary inspirals is
\begin{equation}
\tilde{\Lambda} = \frac{16}{13} \frac{(m_1+12m_2)m_1^4\Lambda_1 + (m_2+12m_1)m_2^4\Lambda_2}{(m_1+m_2)^5}\,,
\end{equation}
where $m_1$ ($m_2$) and $\Lambda_1$ ($\Lambda_2$) are the mass and tidal deformability of the heavier (lighter) component of the binary.
During the inspiral phase of a binary neutron star (BNS) merger, tidal effects enter the gravitational waveform at high post-Newtonian order and are parameterized by the reduced tidal deformability $\tilde{\Lambda}$~\cite{Hinderer:2007mb}. 
Importantly, $\tilde{\Lambda}$ is directly inferred from the signal using Bayesian parameter inference methods with tidal waveform approximants.
In our framework, dark matter modifies stellar structure through changes in the compactness and Love number, thereby shifting $\Lambda(M)$ and consequently $\tilde{\Lambda}$. We assume that the inspiral dynamics remain well described by adiabatic tidal deformations and that no additional dynamical degrees of freedom beyond those encoded in $\Lambda$ are excited. Under this assumption, structural modifications due to DM are captured at leading order through their impact on $\Lambda$.

\paragraph*{\textbf{Applicability of EOS-insensitive relations}:}

Inference of NS radii from GW observations typically employs quasi-universal relations between compactness and tidal deformability calibrated for ordinary neutron stars. 
Systematic deviations in these relations are estimated to be at the few-percent level for current detector sensitivities~\cite{Kashyap:2022wzr}. 
Since $\tilde{\Lambda}$ is directly measured from the inspiral waveform, our primary interpretation focuses on the mass–$\Lambda$ relation. 
Radius-based constraints, which rely on universal relations, are treated cautiously in the presence of non-standard matter components such as dark matter.

\section{Results and Discussions} \label{result}

\subsection{Observational Constraints}

NSs are currently constrained by a combination of GW, X-ray, and radio observations. 
In particular, the BNS merger GW170817~\cite{LIGOScientific:2018cki,Nathanail:2021tay} detected by the LIGO and Virgo detectors, provides constraints on the component masses and on the reduced tidal deformability $\tilde{\Lambda}$. The component masses were inferred to lie within $1.17M_\odot \lesssim (m_1, m_2) \lesssim 1.6M_\odot$ at 90\% credibility~\cite{LIGOScientific:2018hze}. 
Expanding the tidal deformability around a fiducial mass $1.4M_\odot$ yields the commonly quoted constraint
\begin{equation}
\Lambda_{1.4} = 190^{+390}_{-120},
\end{equation}
derived using EOS-insensitive analyses~\cite{DelPozzo:2013ala, LIGOScientific:2018cki}. 
The corresponding radius interval inferred via quasi-universal relations is
\begin{equation}
10.62~{\rm km} < R_{1.4} < 12.83~{\rm km}.
\end{equation}

Independent constraints arise from pulsar observations. 
PSR J0348+0432 and PSR J1614$-$2230 have precisely measured masses of $M = 2.01 \pm 0.04 M_\odot$~\cite{Antoniadis:2013pzd} and $M = 1.908 \pm 0.016 M_\odot$~\cite{Demorest:2010bx, NANOGrav:2017wvv}, respectively. 
NICER measurements of millisecond plsar PSR J0740+6620 provide simultaneous mass-radius constraints; $R = 13.7^{+2.6}_{-1.5} \, \mathrm{km}$ for $M = 2.08 \pm 0.07 M_\odot$ by Miller \textit{et al.}~\cite{Miller:2021qha}, and $R = 12.39^{+1.30}_{-0.98} \, \mathrm{km}$ for $M = 2.072^{+0.067}_{-0.066} M_\odot$ by Riley \textit{et al.}~\cite{Riley:2021pdl}. 
Any viable EOS must support stable configurations consistent with these measurements.

In what follows, we examine whether DANSs constructed within our two-fluid BEC framework can satisfy these constraints and how the presence of DM modifies the mass–$\Lambda$ relation relevant for GW observations.

\subsection{Properties of neutron stars and BEC stars }

We first consider the limiting cases of pure nuclear matter and pure BEC stars. For pure NSs we set $\rho_{\rm DM}=p_{\rm DM}=0$, while for pure BEC stars we set $\rho_{\rm NM}=p_{\rm NM}=0$ in Eqs.~\eqref{eq:tov}.

\begin{figure}[!ht] \label{MRFinal}
\centering
\includegraphics[width=0.5\textwidth]{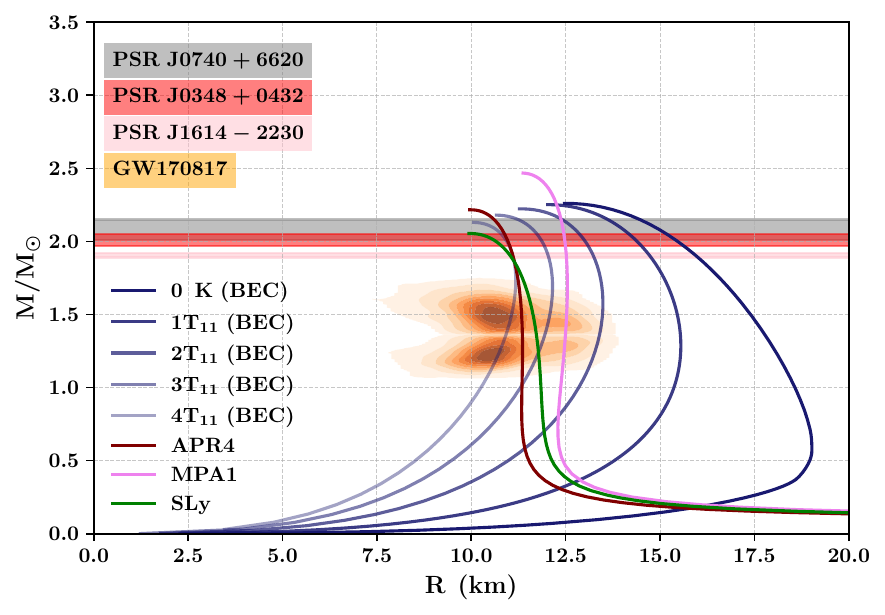}
\caption{Mass-radius relation for pure BEC stars at different temperatures (0-$4 \times 10^{11}$K) and NSs described by APR4, MPA1, and SLy EOSs. The shaded bands represent observational constraints: grey, red, and pink correspond to mass measurements from PSR~J0740+6620, PSR~J0348+0432, and PSR~J1614$-$2230, respectively. The orange contour denotes the 90\% credible region of GW170817 from GW observations.} 
\end{figure}

In Fig.~\ref{MRFinal}, we show the mass as a function of radius for both BEC stars and NSs. 
For BEC stars at $T=0$ ($ 1 T_{11},\,2 T_{11},\,3 T_{11},\,4 T_{11}$) K, we obtain a maximum mass $M = 2.26 \, (2.25,\,2.22,\,2.18,\,2.13) M_{\odot}$,
with a radius $R= 12.46$ ($12.06,11.27,10.66,10.06$) km corresponding to the central density $\rho_c = 2.06 \, (2.08,\,2.24,\,2.36,\,2.51) \times 10^{15}$  
g/cm$^3$. 
The results for zero-temperature BEC stars are consistent with those presented in Ref.~{\cite{Chavanis:2011cz}}. 
For finite temperature cases, we find that stellar equilibria occur at slightly reduced masses and radii, though these changes are relatively minor. 
This behavior aligns with earlier studies of static BEC stars at finite temperatures using alternative EOS, as reported in Ref.~{\cite{Latifah:2014ima}}. 
Figure~\ref{MRFinal} indicates that increasing temperature leads to a softer EOS, which can be attributed to thermal fluctuations within the star. 
However, in Ref.~\cite{Latifah:2014ima}, the use of a different EOS formulation resulted in a stiffer EOS at higher temperatures, leading to distinct trends in the mass-radius ($M$-$R$) relationship. 
These differences stem from varying theoretical approaches, producing contrasting temperature-dependent EOS behaviors and making a direct comparison between their findings and our results inappropriate.
Figure~\ref{MRFinal} also shows the well-known $M$-$R$ curves for the realistic NS EOSs. 
All three EOSs satisfy the maximum mass constraints from the pulsar observations.

\begin{figure}[!ht] \label{lambdaFinal}
\centering
\includegraphics[width=0.5\textwidth]{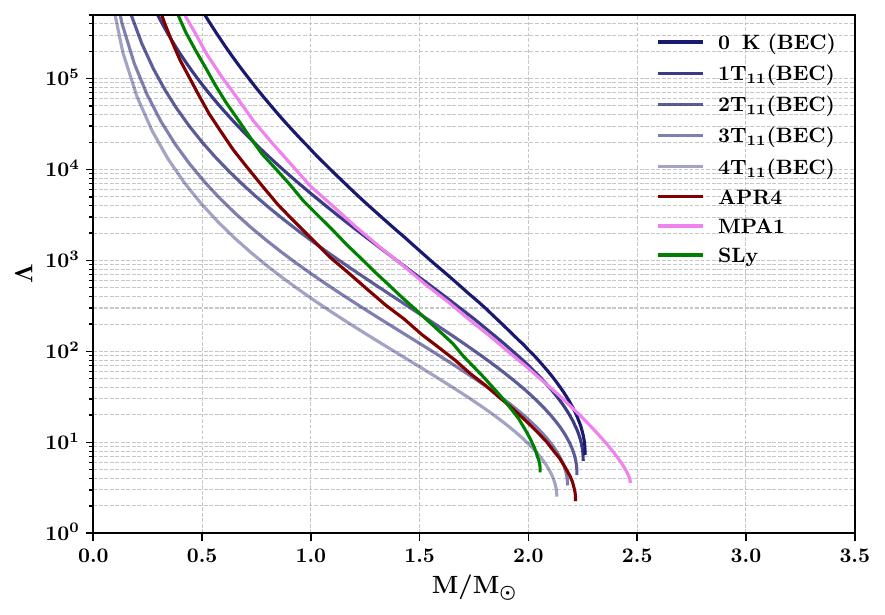}
\caption{Dimensionless tidal deformability $\Lambda(M)$ as a function of mass for pure BEC stars and neutron stars described by the EOSs: APR4, MPA1, and SLy.}
\end{figure}

The parameter that is directly measurable by GW observations of a BNS inspiral is
proportional to the tidal deformability $\Lambda$, shown for BEC stars and NSs in Fig.~\ref{lambdaFinal}. As a general behavior in these plots, it can be seen that tidal deformability is a decreasing function of mass. 
For completeness, we compare the features of a canonical $M = 1.4 M_{\odot}$ BEC star and NS models in the Table~\ref{table:1}.

\begin{table}[h]
    \centering
    \setlength{\tabcolsep}{10pt}
    \begin{tabular}{|c|c|c|}
        \hline
        $T$ ($T_{11}$ K) & $R$ (km) & $\Lambda$ \\ \hline
        0 & 17.6 & 2079 \\ \hline
        1 & 15.5 & 981 \\ \hline
        2 & 13.4 & 372 \\ \hline
        3 & 12.0 & 172 \\ \hline
        4 & 10.9 & 95 \\ \hline
        APR4 & 11.3 & 259 \\ \hline
        MPA1 & 12.5 & 398 \\ \hline
        SLy & 11.7 & 276 \\ \hline
    \end{tabular}
    \caption{Properties of $M = 1.4 M_{\odot}$ pure BEC stars and neutron star models.}
    \label{table:1}
\end{table}

\subsection{Dark matter admixed neutron stars}

\begin{figure*}[!ht]
    \centering
    \includegraphics[width=\linewidth]{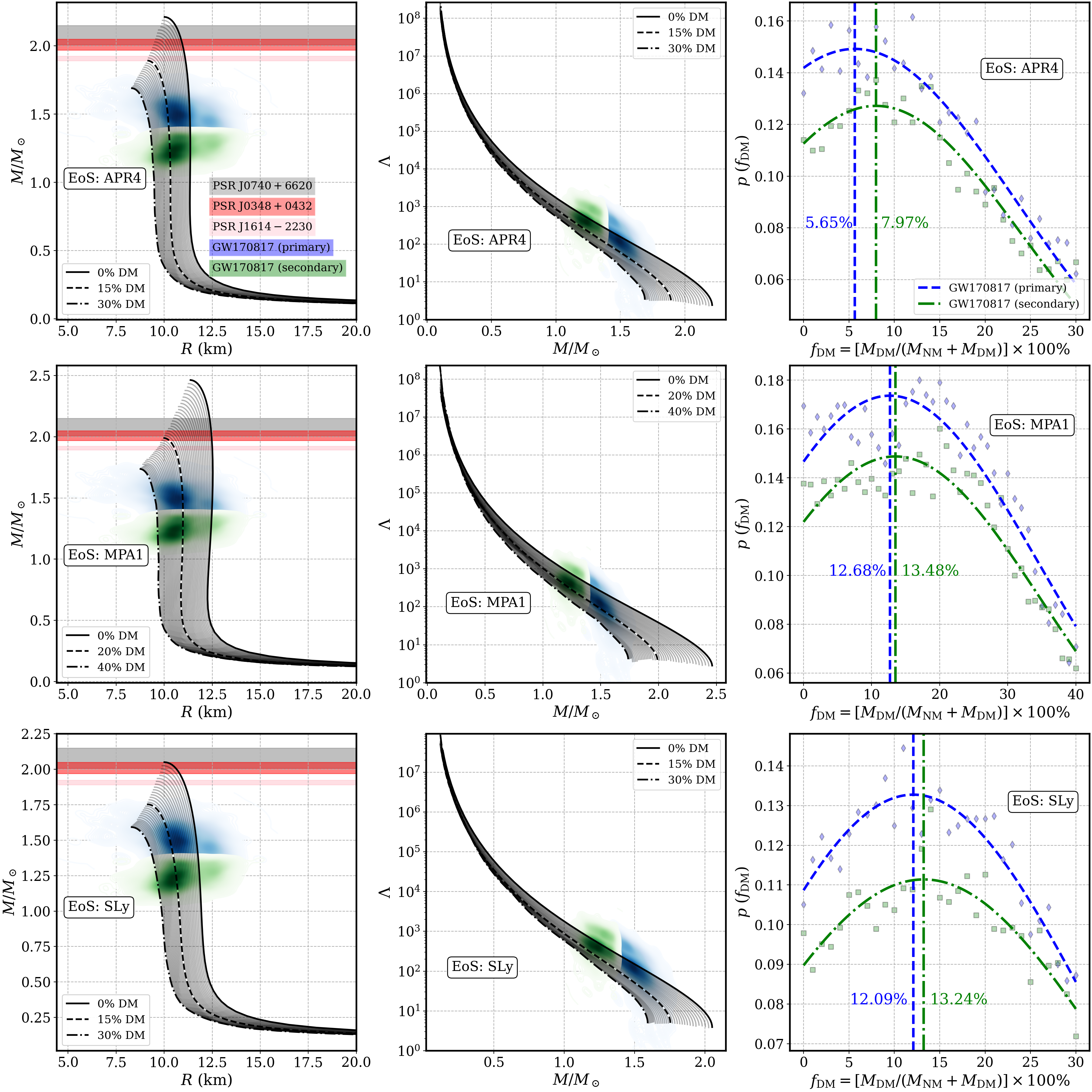}
    \caption{Mass-radius relations (first column), variation of the dimensionless tidal deformability $\Lambda$ (middle column), and the distribution of probabilities $p(f_{\rm DM})$ (equation~\eqref{eq:p_f}) of a dark matter fraction $f_{\rm DM}$ being present in the NSs observed in the GW170817 event (third column) for the EOSs APR4, MPA1, and SLy. The constraint patches from pulsar observations and the GW event are the same as in Fig.~\ref{MRFinal}, but posteriors for the two stars in GW170817 are plotted in different colors in this plot, and the same color choice is maintained in plotting $p(f_{\rm DM})$ on the third column. In the first two columns, the DM fraction is varied from 0 to 30\% for APR4 and SLy, and up to 40\% for MPA1, with intermediate curves shown in 1\% increments. $M$-$R$ and $M$-$\Lambda$ curves with the highest dark matter and half that value are highlighted for reference. Each curve on the $M$-$R$ plots corresponds to a $p(f_{\rm DM})$ value on the third column, and the distribution is estimated by a Gaussian fit, for which the median values are reported for both stars.}
    \label{fig:MRlambdapf}
\end{figure*}

We now consider two-fluid configurations with varying dark matter fraction $f_{\rm DM}$. 
The first two columns of Fig.~\ref{fig:MRlambdapf} illustrate the influence of DM on NSs modeled with the EOSs APR4, MPA1, and SLy. 
As $f_{\rm DM}$ increases, both the maximum mass and the corresponding radius of the NSs decrease noticeably. 
Consistent with previous studies, the presence of a DM core reduces the maximum stable mass and the corresponding minimum radius, highlighting the significant impact of DM on the structure of an NS~\cite{Ellis:2018bkr}. 
On these plots, we highlight the constraints obtained from pulsar observations as discussed before, and also the constraints on the properties of the NSs observed in GW170817. 
For APR4, we see that an NS can contain $\sim 10\%$ DM in its core while still conforming to the $\sim 2M_\odot$ pulsar observations. 
For MPA1, this tolerance increases to $\sim 20\%$, while SLy can sustain only up to $\sim 2\%$ DM to keep this bound satisfied. 

We observe that the tidal deformability consistently decreases as the mass of the DM core increases, regardless of the EOS under consideration. 
As indicated in Eq.~\eqref{Lambda}, $\Lambda \propto (R/M)^5$. 
Consequently, its minimum value is associated with the maximum mass and minimum radius of the DANS.

Assuming that there is no DM present, the $M$-$R$ curves produced by all three EOSs considered in this work predict higher masses and radii than their most probable region of the GW170817 posteriors for both the components NSs of the binary. 
However, entertaining the possibility of the presence of DM in their core opens up the prospect of each of them being a better candidate nuclear matter EOS for the two compact objects observed in GW170817, with different DM percentages. 

We quantify this by defining the conditional probability distribution $p(f_{\rm DM})$ that one of the NSs in this GW event contained a fraction $f_{\rm DM}$ of DM in its core, given the posterior samples released by the LIGO-Virgo collaboration, as
\begin{equation}\label{eq:p_f}
    p(f_{\rm DM})=\int p^0(M,R)\delta(M(R)-M(R;f_{\rm DM}))\,\dd M\,\dd R\,,
\end{equation}
where $p^0(M,R)$ is the probability of the NSs having a mass $M$ and radius $R$, and can be calculated from the posterior samples. 
$M(R;f_{\rm DM})$ dentoes the $M$-$R$ relation with a DM fraction $f_{\rm DM}$, under the two-fluid scenario adopted here. 
Equation~\eqref{eq:p_f} is based on the work on constraining nucleonic EOS by Kashyap \textit{et al.}~\cite{Kashyap:2025cpd}. 
We used the EOS-insensitive posterior of GW170817 component objects~\cite{LIGOScientific:2018cki} to calculate $p(f_{\rm DM})$, and the results are shown in the third column of Fig.~\ref{fig:MRlambdapf}. 
$p(f_{\rm DM})$ is calculated for discrete values of the DM fraction, and the resulting distributions are fitted with a Gaussian function to estimate the most probable values. 
One can interpret how the two-fluid DANS scenario predicts the DM content of the component NSs in the GW170817 event from these plots. 
For example, if APR4 is the true nuclear matter EOS describing NSs in the Universe, and the component objects in this event were indeed DANSs, then the most probable DM content in their cores is $\sim 5.65\%$ and $\sim 7.97\%$, respectively. 
These predicted DM fractions are much higher (above $12\%$) for the other two EOSs, as seen from the figures. 
The possibility of the presence of DM thus introduces complications in constraining the true EOS for nuclear matter, as the observed properties of the NSs in GW events may be consistent with multiple EOSs with different DM fractions.

\begin{figure*}[!ht]
    \centering
    \subfigure[]{\includegraphics[width=\linewidth]{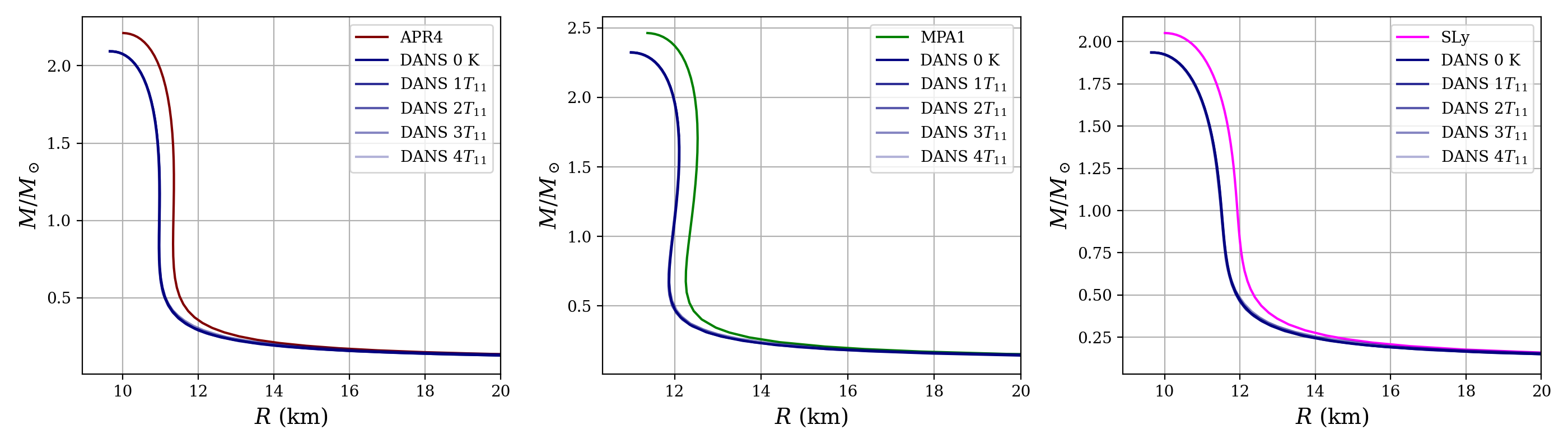}}\\
    \subfigure[]{\includegraphics[width=\linewidth]{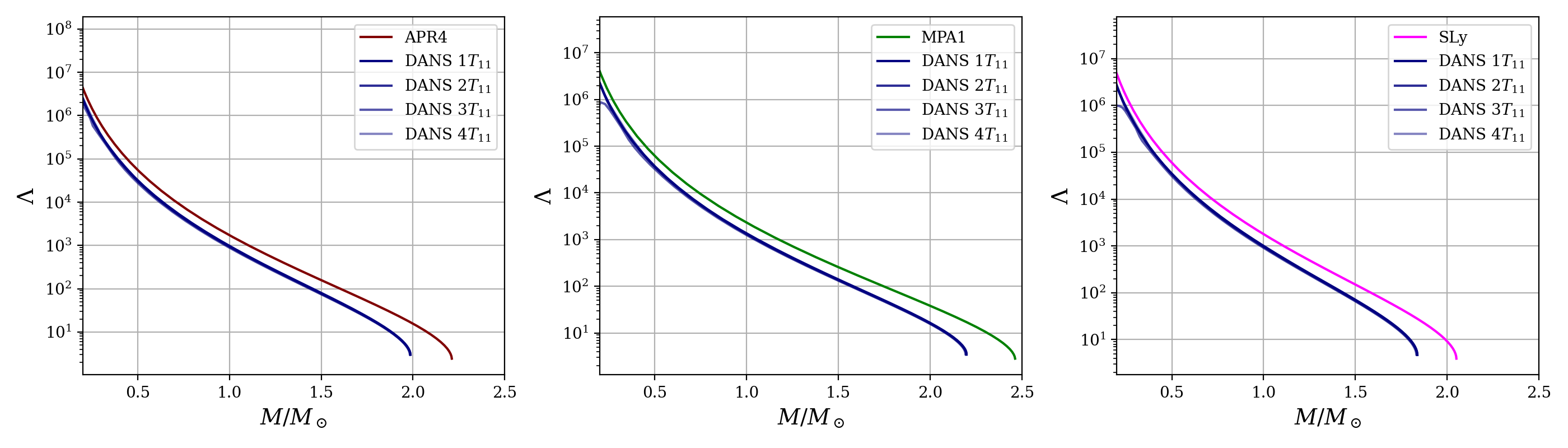}}
    \caption{Effects of finite temperature on DANS with fixed $f_{\rm DM}=5\%$. $T_{11}=10^{11}$ K.}
    \label{fig:m-r-lambda-with-temp}
\end{figure*}

To investigate the impact of the temperature of the BEC DM on the properties of a DANS, we fix the DM content to 5\% and vary the DM temperature from 0 K to 4 $T_{11}$. 
Fig.~\ref{fig:m-r-lambda-with-temp} illustrates the resulting $M$-$R$ (top panels) and $M$-$\Lambda$ (bottom panels) relations for the three EOSs APR4, MPA1, and SLy. 
Whereas the inclusion of $5\%$ dark matter reduces the maximum mass of the neutron star from $2.21 M_{\odot}$ to $2.09 M_{\odot}$ at 0 K using the APR4 EOS, we observe that the stability criteria are largely unchanged even with the temperature reaching as high as $4\times 10^{11}$ K. 
The influence of finite temperature on the DANS system remains negligible for all three EOSs, as is seen from this figure. 
For pure BEC stars, Fig.~\ref{MRFinal} demonstrates the significant effect of temperature on the stability criteria and thus the $M$-$R$ relations. 
However, in the case of DANSs, Fig.~\ref{fig:m-r-lambda-with-temp} shows that this effect is strongly suppressed. 
This behavior can be attributed to the fact that whereas the presence of BEC DM may have substantial effects on the stars, the stability and tidal properties of the stars are primarily governed by the mass content of the two fluids. 
We conclude, therefore, that in the context of GW observations, the temperature of the BEC DM is largely inconsequential. 

\section{Conclusion} \label{conclusion}

In this work, we have investigated neutron stars admixed with dark matter in the form of a finite-temperature Bose–Einstein condensate within a general relativistic two-fluid framework. 
The nuclear and dark sectors are assumed to interact only gravitationally, and equilibrium configurations are obtained by solving the coupled TOV equations. 
For the nuclear matter EOS, we consider APR4, MPA1, and SLy, all consistent with current $2M_\odot$ maximum mass constraints for NSs from pulsar observations.

We find that the presence of a DM core generically reduces the maximum mass, radius, and tidal deformability of neutron stars. 
This behavior is largely independent of the choice of nuclear EOS. 
For a pure neutron star described by the APR4 EOS, the maximum mass is $2.21\, M_{\odot}$ with a radius of $10.02$ km. 
When $5\%$ DM is included in the core, the maximum mass decreases to $2.09\, M_{\odot}$, and the corresponding radius reduces to $9.68$ km. 
Since the tidal deformability scales approximately as $\Lambda \propto (R/M)^5$, even modest changes in compactness lead to appreciable shifts in $\Lambda$. 
These structural modifications directly affect the mass–$\Lambda$ relation relevant for GW observations.

We illustrate how the two-fluid DANS model can be employed to estimate the DM content in the NSs involved in the GW170817 event. 
Assuming that APR4 is the correct nuclear matter EOS in nature and that the component objects in this BNS merger were indeed DANSs, the most probable DM fractions in their cores are estimated to be approximately $5.65\%$ and $7.97\%$, respectively. 
For the other two EOSs considered, the inferred DM content exceeds $12\%$. 
The tidal deformability consistently decreases with an increase in the mass of the DM core, irrespective of the EOS being considered.

We stress that this interpretation is conditional. 
It assumes that (i) the inspiral waveform is adequately described by adiabatic tidal effects parameterized by $\Lambda$, and (ii) dark matter modifies the GW signal predominantly through changes in the stellar structure and tidal deformability. 
Within this structural framework, DM introduces an additional degeneracy in EOS inference: multiple nuclear EOSs can reproduce the same observed tidal parameters if different DM fractions are allowed.

We have also examined the effect of finite temperature in the BEC sector. 
While temperature significantly affects pure BEC stars, its impact on DANSs is strongly suppressed when the DM fraction is small. 
Even at temperatures as high as $4\times10^{11}$ K, the mass–radius and mass–$\Lambda$ relations of DANSs remain nearly identical to the zero-temperature case for the parameter space explored here. 
Thus, in the context of current gravitational-wave observations of inspiralling binaries, finite-temperature corrections to the BEC dark matter EOS appear to be negligible.

Several extensions of this work remain open. 
We have assumed both NM and DM to be isotropic perfect fluids. 
In realistic scenarios, anisotropies or additional microphysical interactions may arise, potentially affecting both equilibrium structure and tidal response~\cite{Mahapatra:2024ywx, Liu:2025cwy}. 
Incorporating such effects, as well as exploring non-adiabatic or dynamical DM degrees of freedom, would provide a more complete picture. 
Furthermore, future GW observations with improved precision may help disentangle degeneracies between nuclear EOS stiffness and possible DM contributions. 

In summary, our study aims to make a significant and fundamental contribution to the understanding of DANSs. Since DM does not interact electromagnetically, and GW experiments cannot distinguish between different forms of matter, the resulting tidal effects depend solely on the compactness of the object. Upcoming data from LIGO and future GW observatories such as LISA \cite{amaro2017laser}, the Einstein Telescope \cite{maggiore2020science}, and the Cosmic Explorer \cite{reitze2019cosmic} will provide valuable opportunities to identify NSs containing subfractions of DM or similar exotic particles. This approach not only aligns more closely with observational data but also enables more sophisticated modeling of exotic matter and GW predictions, advancing our understanding of compact objects in the universe.

\section{Acknowledgments} \label{Acknowledgments}

AG thanks Rahul Kashyap for the detailed discussion on the dark matter fraction constraints. SM thanks P. Ajith and Prasad R. for their useful comments. PSA would like to thank IUCAA for providing the opportunity and resources to carry out this research project. The authors acknowledge the computational resources provided by the Sarathi cluster at IUCAA.

\bibliography{reference.bib}

\end{document}